\documentclass[a4paper]{jpconf}
\usepackage{graphicx,amssymb}
\begin{document}
\title{Testing common classical LTE and NLTE
model atmosphere and line-formation codes for quantitative spectroscopy
of early-type stars}

\author{Norbert Przybilla$^1$, Maria-Fernanda Nieva$^{1,2}$ and Keith Butler$^3$}

\address{$^1$ Dr. Karl Remeis-Sternwarte \& ECAP, Sternwartstr. 7, D-96049 Bamberg, Germany}
\address{$^2$ Max-Planck-Institut f\"ur Astrophysik, Karl-Schwarzschild-Str.~1, D-85741 Garching, Germany}
\address{$^3$ Universit\"atssternwarte, Scheinerstr. 1, D-86179 M\"unchen, Germany}

\ead{norbert.przybilla@sternwarte.uni-erlangen.de}

\begin{abstract}
It is generally accepted that the atmospheres of cool/lukewarm stars of spectral
types A and later are described well by LTE model atmospheres, while
the O-type stars require a detailed treatment of NLTE effects. 
Here model atmosphere structures, 
spectral energy distributions and synthetic spectra computed with 
{\sc Atlas9/Synthe} and {\sc
Tlusty/Synspec}, and results from a hybrid method combining LTE
atmospheres and NLTE line-formation with {\sc
Detail/Surface} are compared. Their ability to reproduce 
observations for effective temperatures 
between 15\,000 and 35\,000\,K are verified. Strengths and weaknesses
of the different approaches are identified. Recommendations are made as to how to
improve the models in order to derive unbiased stellar parameters and
chemical abundances in future applications, with special 
emphasis~on~Gaia~science.
\end{abstract}

\section{Introduction}
Quantitative stellar spectroscopy employs model atmospheres that are
based on many approxi\-mations and simplifications. One of the fundamental
decisions to be made for the model construction concerns the description 
of the thermodynamic state of the atmospheric plasma. Either local thermodynamic
equilibrium (LTE) may be assumed, or deviations from it may be allowed
for (NLTE), i.e. taking the mutual interaction of radiation and
matter into account.

It is generally accepted that the atmospheres of cool/lukewarm stars
of spectral types A and later are described well by LTE model
atmospheres, while O stars require a detailed treatment of 
NLTE effects due to the high energy density of their radiation field.
Both approaches are being followed in the literature for analyses of
B-type stars at present, with variable success.

The present work tests the predictive power of common LTE and NLTE
model atmosphere and line-formation codes in this transition region, 
concentrating on main-sequence stars of effective temperatures
between 15\,000 and 35\,000\,K (corresponding to spectral types of 
about B4 to O9). Three approaches are investigated:
pure LTE as implemented in the {\sc Atlas9/Synthe} suite of codes
\cite{k93b,ka81}, pure NLTE with {\sc Tlusty/Synspec} \cite{hl95},
and a hybrid approach based on LTE atmospheres ({\sc Atlas9}) and
subsequent NLTE line-formation computations with {\sc Detail/Surface}
\cite{g81,bg85}. We deviate from the usual `model~vs.~model'
comparison -- which may be misleading as {\it a priori} it is not clear 
as to which model is more realistic -- by testing 
observational constraints wherever possible.

\begin{figure}
\begin{center}
\includegraphics[width=.62\linewidth]{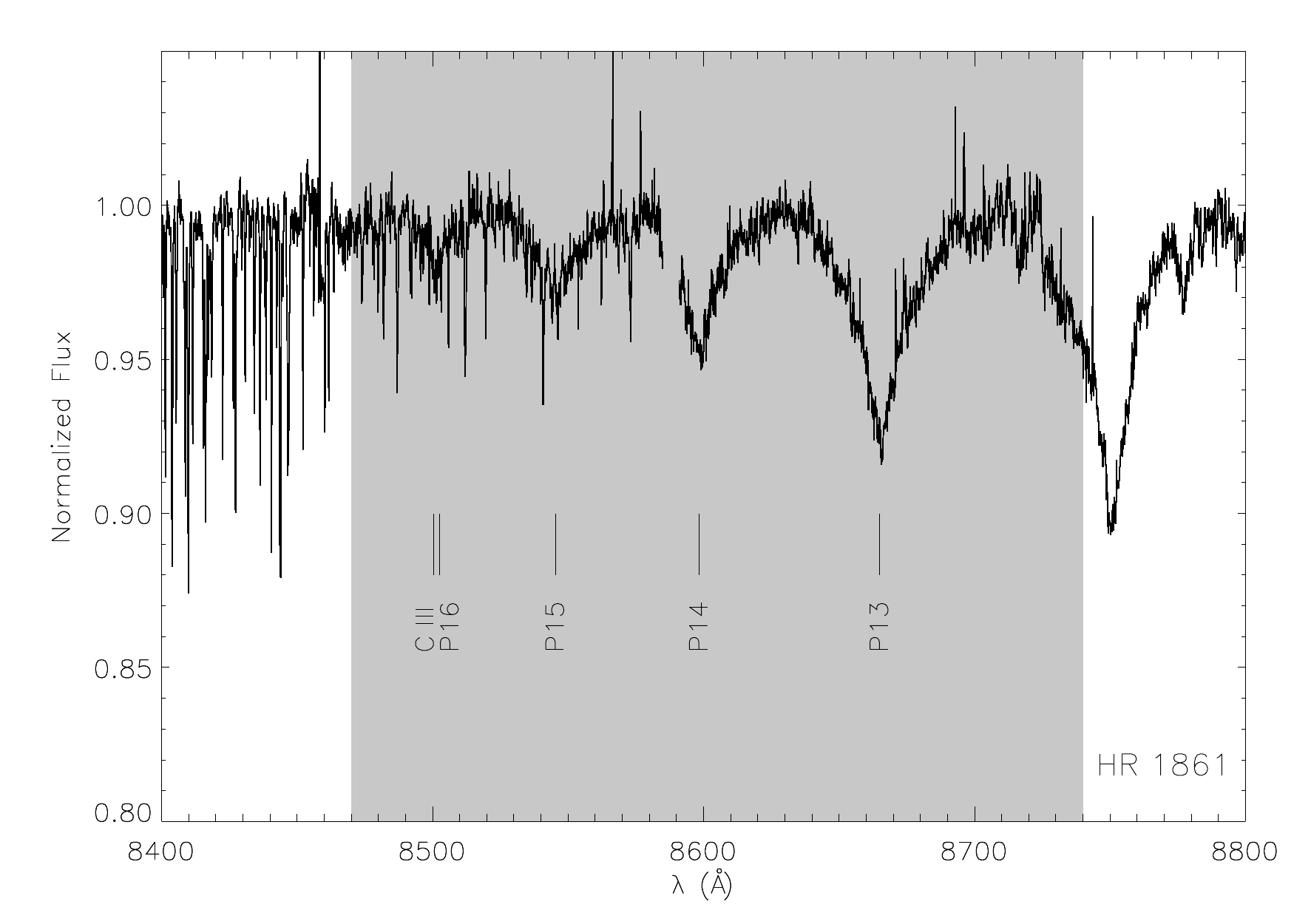}
\end{center}
\vspace{-5mm}
\caption{Near-IR spectrum of HR\,1861 (B1\,IV), observed with {\sc
Foces} on the Calar Alto 2.2m telescope. The most important
lines in the Gaia RVS spectral range (grey region) are identified.}
\label{gaiarvs}
\end{figure}

\section{The Gaia perspective for investigations of OB-type stars}
The legacy of the Gaia mission in terms of OB-type stars will mainly
lie in the determination of highly precise parallaxes, proper motions
and broad-band space photometry. In contrast to the cooler stars,
little information besides radial and projected 
rotational velocities will be derivable from the
spectra obtained with the on-board radial-velocity spectrometer (RVS).
The reason for this is the scarcity of spectral lines in the
wavelength range of the RVS, which essentially comprises the 
Paschen series members P$_{13}$ to P$_{16}$ in this kind of star, 
see Fig.~\ref{gaiarvs}. 
The determination of accurate and precise stellar parameters and
elemental abundances will therefore have to rely on ground-based
observations, e.g. within the Gaia-ESO survey \cite{b11}.

\section{Models: assumptions \& implementations}
Quantitative spectroscopy of early-type main-sequence stars and giants 
in the effective
tem\-perature range of interest, 15\,000\,$\le$\,$T_\mathrm{eff}$\,$\le$\,35\,000\,K, 
can rely on {\em classical model atmospheres}\footnote{More luminous
stars, either mid/early O-type stars or OB-supergiants close to the
Eddington limit, require consideration of their stellar winds, 
i.e.~hydrodynamic model atmospheres (and spherical geometry), see
e.g.~\cite{g11}.}. 
Basic assumptions for the model construction are chemical homogeneity, 
stationarity, plane-parallel geometry and 
hydrostatic and radiative equilibrium, while the thermodynamic state
of the atmospheric plasma can be determined in 
LTE or NLTE. The former case is conceptually less realistic but
attractive since typical computing times per model are a few seconds on modern PCs, while NLTE models may require many CPU
hours. A third approach, so-called hybrid NLTE modelling, can combine
the strengths of both types of models, performing sophisticated NLTE
line-formation computations on LTE atmospheres, at a fraction of the
computing cost of full NLTE models. This is particularly attractive
in those cases where the atmospheric structure is close to LTE as for
the parameter space investigated here which we will show in the following. 

\begin{table}[ht!]
\caption{\label{atoms}Model atoms used with our {\sc Detail/Surface} computations.}
\begin{center}
\begin{tabular}{ll@{\hspace{12mm}}ll@{\hspace{12mm}}ll@{\hspace{12mm}}ll}
\br
Ion & Ref. & Ion & Ref. & Ion & Ref. & Ion & Ref.\\
\mr
H\,{\sc i}     & \cite{pb04}         & C\,{\sc ii/iii}    & \cite{np06,np08} & 
O\,{\sc i/ii}  & \cite{pbbkv00,bb88} & Si\,{\sc iii/iv}   & \cite{bb90}\\
He\,{\sc i/ii} & \cite{p05}          & N\,{\sc ii}        & \cite{pb01}      & 
Mg\,{\sc ii}   & \cite{pbbk01}       & Fe\,{\sc ii/iii}   & \cite{b98,mba06}\\
\br
\end{tabular}
\end{center}
\end{table}

For our test we have chosen results from the most common among the many 
available model atmosphere codes: on the one hand line-blanketed LTE 
atmospheres and synthetic spectra computed with the 
{\sc Atlas9/Synthe} suite of codes \cite{k93b,ka81} 
and on the other line-blanketed NLTE models calculated with the 
{\sc Tlusty/Synspec} package \cite{hl95}. Model structures, fluxes and
detailed
synthetic spectra were adopted from published grids, Castelli's ODFNEW
models \cite{ck03}, the Padova spectral library~\cite{mscz05},
and the OSTAR2002 and BSTAR2006 grids \cite{lh03,lh07}.

Our own computations rely on {\sc Atlas9} atmospheres and NLTE
line-formation computations with updated and extended versions of {\sc
Detail/Surface} \cite{g81,bg85}. The former solves the coupled
statistical equilibrium and radiative transfer equations, employing the
accelerated lambda iteration scheme of \cite{rh91}, while the
latter computes the synthetic spectra, using refined line-broadening
theories. Model atoms according to Table~\ref{atoms} were employed.
More details on our hybrid NLTE approach (labeled `{\sc Ads}' henceforth) 
can be found e.g.~in \cite{np07}.

\begin{figure}
\begin{center}
\includegraphics[width=.79\linewidth]{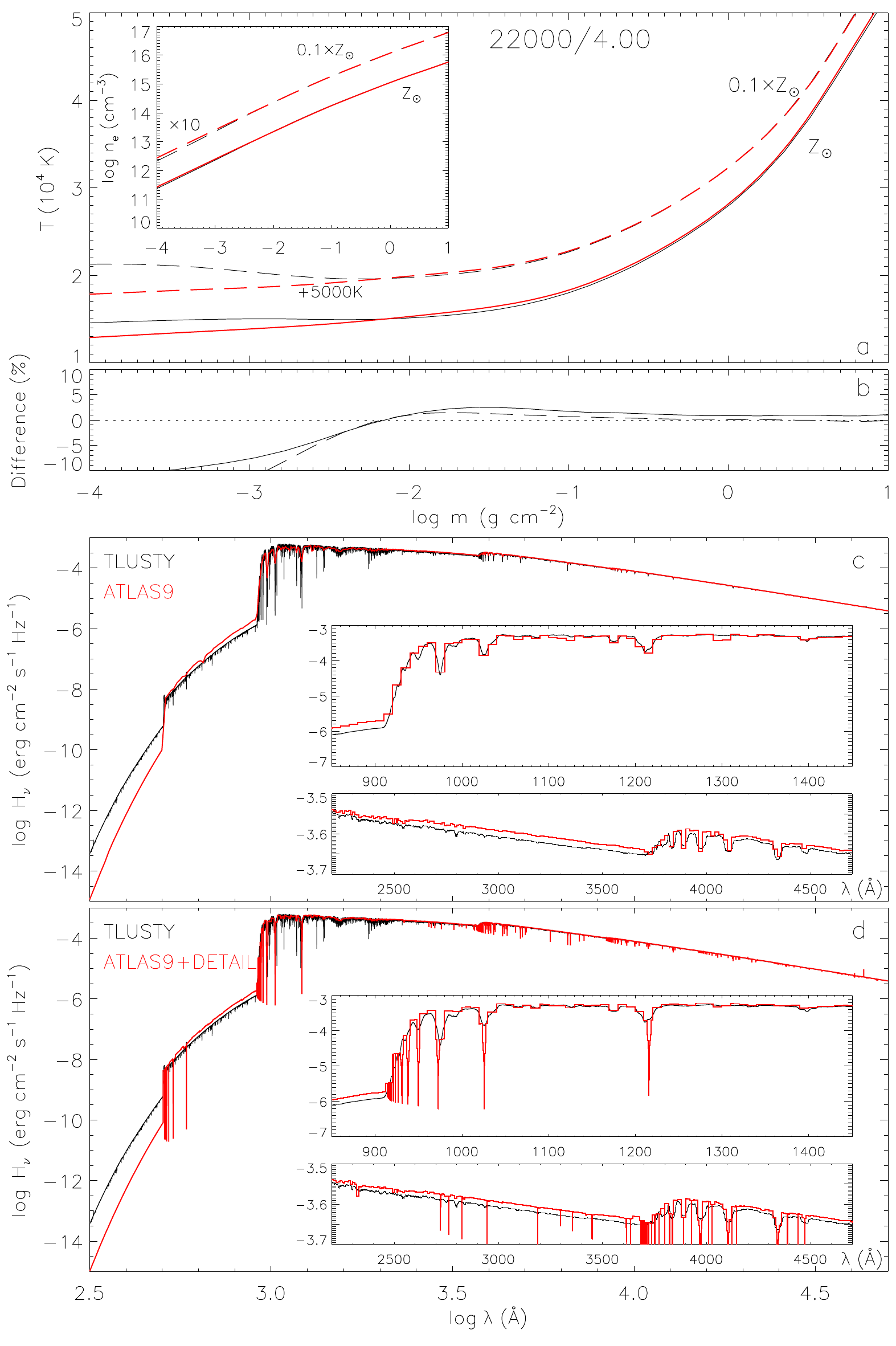}
\end{center}
\vspace{-5mm}
\caption{Comparison of model atmosphere structures and resulting
spectral energy distributions. Panel (a) presents a comparison of
{\sc Atlas9} (red) and {\sc Tlusty} (black)
temperature structures as a function of mass scale, 
for models at $T_\mathrm{eff}$\,=\,22\,000\,K,
$\log g$\,=\,4.00, $\xi$\,=\,2\,km/s at solar (full lines)
and 1/10th solar metallicity (dashed lines, shifted by $+$5000\,K). 
Electron density structures are shown in the inset. Panel (b) displays the 
resulting differences between the {\sc Atlas9} and {\sc Tlusty} 
temperature structures (in percent). Panel (c) and (d) compare the
{\sc Tlusty} with {\sc Atlas9} and {\sc Atlas9+Detail} Eddington fluxes,
respectively. The insets highlight the regions around the Lyman and
the Balmer jumps. Colour coding according to the legends.}
\label{structures}
\end{figure}

\section{Model tests}
In order to facilitate the use of existing grids for the model
comparisons, a two-step strategy has to be followed. In the first
step, we compare our {\em tailored} {\sc Ads} modelling with observation for
strategic points in the investigated parameter space to test its
ability to reproduce observation. Then, in the second step, {\sc
Ads} models for {\em generic} parameters are compared 
to the {\sc Tlusty/Synspec} and {\sc Atlas9/Synthe} models at a grid
point closest to the real star's parameters, thus facilitating a 
meaningful assessment of the predictive power of all these 
modelling implementations. While the synthetic spectra were convolved
with the same rotational and macroturbulent broadening profiles 
as for the comparison with observation, the
Gaussian (`instrumental') profile had to be adjusted to account for 
the lower resolution of the Padova grid. Due to space
restrictions we will discuss the case of
the well-studied standard star
$\gamma$\,Peg (B2\,IV) in detail in the following, and comment on 
more general aspects
of the comparisons~briefly\footnote{Our observational sample 
comprises a total of 29
apparently slowly-rotating B3 to B0 dwarf and giant stars, for
which spectra of high signal-to-noise ratio (S/N\,$\sim$\,250--800), 
high resolution ($R$\,=\,$\Delta\lambda/\lambda$\,$\simeq$40--48\,000) and wide
wavelength coverage were obtained using state-of-the-art Echelle
spectrographs on 2m-class telescopes, see \cite{npi11} for a brief
overview.  For this sample we have shown that combining a sophisticated 
analysis technique with our {\sc Ads} modelling allows  
practically the entire observed spectra in the optical for all these objects
\cite{pnb08,nsd11,np11} to be reproduced to similar or even better 
quality as shown for $\gamma$\,Peg in the following.}.

For $\gamma$\,Peg (HD\,886) we derived
$T_\mathrm{eff}$\,=\,22\,000\,K, a surface gravity of $\log
g$\,=\,3.95 (cgs units), micro\-turbulence $\xi$\,=\,2\,km/s, solar
helium abundance and metal abundances typical for the normal early
B-type star population in the solar neighbourhood \cite{pnb08,np11}.
The nearest parameter combination covered by the published grids
coincides with these, except that $\log g$\,=\,4.00 and 
solar abundances according to \cite{gs98}.
We have adopted the generic grid parameters nevertheless 
for the {\sc Ads} model in the second step of the comparison in order
to avoid any systematic bias.\\[-8mm]  

\paragraph{Model atmosphere structures and spectral energy distributions.}
We begin testing the models with a comparison of
atmospheric structures in Fig.~\ref{structures}. The temperature
structures of the LTE and NLTE models agree very well, except for the
outermost layers. Differences amount to up to 1--2\% for all depths
relevant for the formation of the spectral lines and continua 
from the far-UV through the optical to the near-IR. Interestingly, 
the differences are
smaller for the low-metallicity comparison at 1/10th of the solar value,
despite the fact that NLTE effects should be more pronounced because of increased
photon mean-free paths and a harder radiation field. 
A possible explanation of this unexpected finding could be the
differences in the metal line opacities (which matter more at higher
metallicity ). Many more elements are considered in the {\sc Atlas9} 
than in the {\sc Tlusty}
model, resulting in a stronger line-blanketing effect, exactly as
indicated in Fig.~\ref{structures}\,(a). Very good agreement is found 
also for LTE and NLTE density structures.
Overall, close agreement of the LTE and NLTE structures is found 
over the entire parameter range considered, for the depths relevant for
the formation of the observable line spectra and continua,
at $-$2\,$\lesssim$\,$\log m$\,$\lesssim$\,$-$0.3.

As temperature and density stratifications are not directly
observable, we verified the effects on line profiles of
hydrogen, helium and metal lines by performing line-formation
computations with {\sc Detail \& Surface} on {\sc Atlas9} and {\sc
Tlusty} atmospheres. Indeed, only small differences are found for the
resulting profiles, which are irrelevant for the stellar parameter and
elemental abundance determination, see e.g.~Fig.~10 of \cite{np07}. 
Maximum effects on metal abundances derived from a few individual lines 
are about 0.05\,dex, and less for mean elemental abundances.

\begin{figure}
\begin{center}
\includegraphics[width=.66\linewidth]{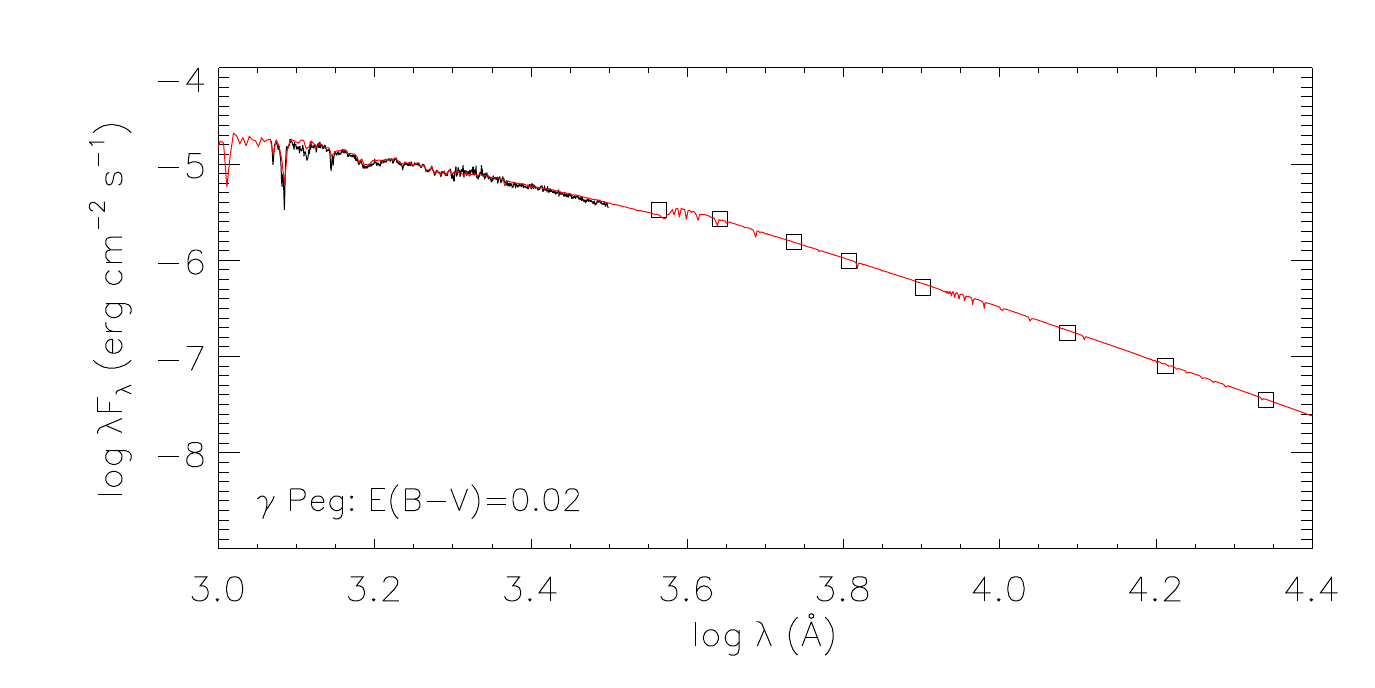}
\end{center}
\vspace{-5mm}
\caption{Comparison between our computed SED (red line) and observation
for $\gamma$\,Peg (B2\,IV):
IUE spectro\-photometry (black line) and UBVRIJHK photometry (squares).
}
\label{gpSED}
\end{figure}

Overall good agreement is also found for the LTE and NLTE spectral
energy distributions (SEDs), as exemplified in Fig.~\ref{structures} (c) and
(d). The differences in the continua are because of the temperature
differences at the formation depths. Significant discrepancies occur only in
the Lyman and helium continua, but it is difficult to test the models
in this respect as interstellar absorption prevents observations of
the extreme-UV for almost all early-type stars. The model differences
in the extreme-UV vary with $T_\mathrm{eff}$. Very
good agreement over the entire spectral range is achieved at higher
$T_\mathrm{eff}$, see Fig.~9 of \cite{np07}. Our models can reproduce
observed SEDs and therefore the global stellar energy output very well, if 
the observations are corrected for the
appropriate amount of interstellar reddening. An example, for
$\gamma$\,Peg, is shown in Fig.~\ref{gpSED}.

\begin{figure}
\begin{center}
\includegraphics[width=.9\linewidth]{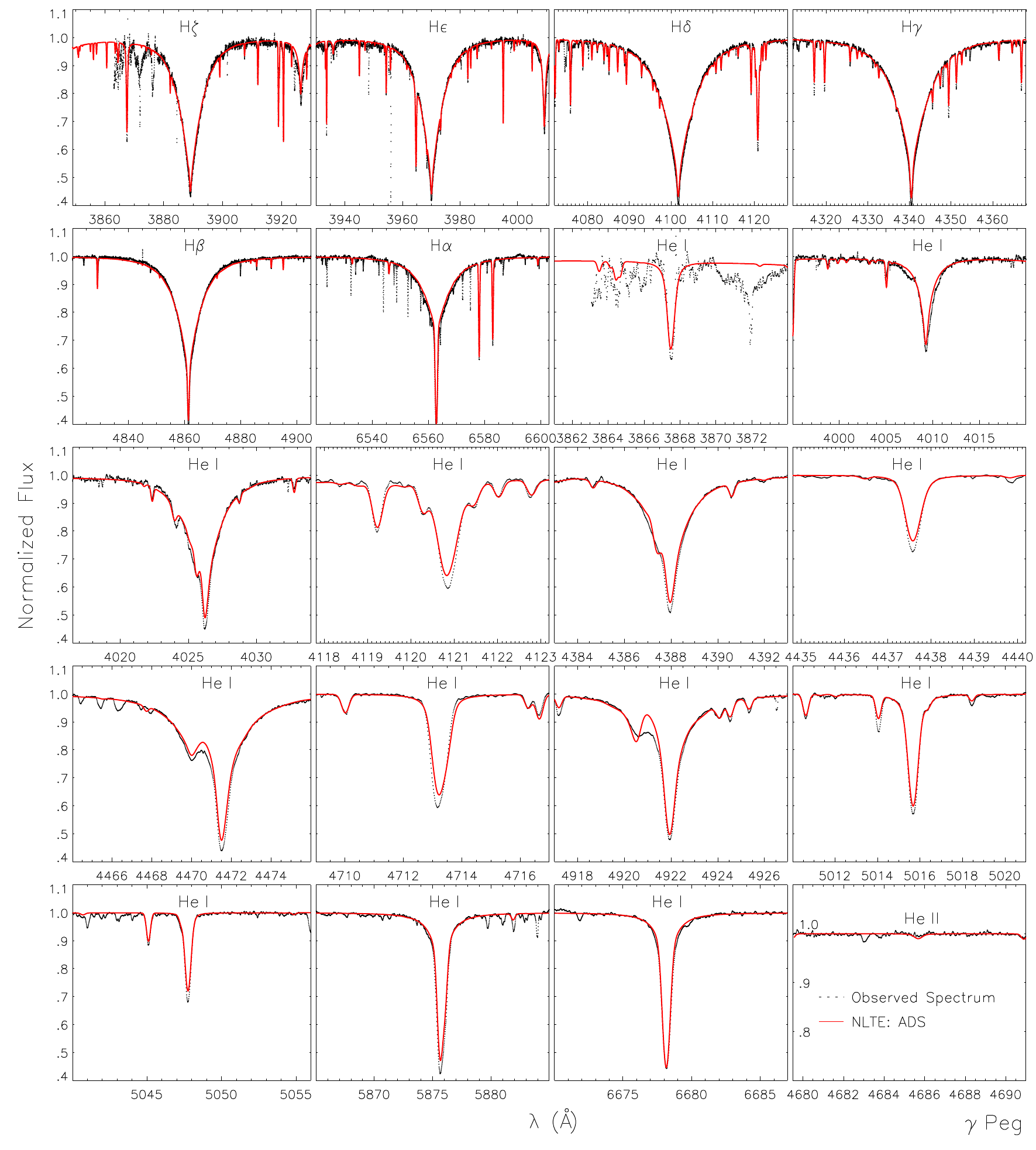}
\end{center}
\vspace{-5mm}
\caption{Comparison of strategic H\,{\sc i} and He\,{\sc i/ii} in
the observed spectrum of $\gamma$\,Peg (black dots) with our global 
best-fit {\sc Ads} model (red line).} 
\label{HHEo}
\end{figure}

We conclude that {\it LTE and NLTE model atmospheres are essentially
equivalent for dwarf and giant stars over the range
15\,000\,$\le$\,$T_\mathrm{eff}$\,$\le$\,35\,000\,K}, for most practical
applications.\\[-8mm]

\paragraph{Line spectra.} Resolved line spectra are the most valuable
observables as they give information about the temperature and 
density stratifications within the stars' atmospheres. They are
also well-suited for the determination of stellar parameters and
elemental abundances.\\[1mm]
\noindent \underline{H \& He}: Hydrogen and helium provide the deepest features in
the spectra of early B-type stars, thus sampling the widest extent of
the atmospheric stratification. Our overall best-fit {\sc Ads} model is
compared to observed profiles for many diagnostic lines of $\gamma$\,Peg 
in Fig.~\ref{HHEo}. We note explicitly, that {\it one} synthetic
spectrum is displayed in all the individual panels -- also for all following
comparisons -- and not best fits to the individual lines.
Overall, an excellent match of observation and theory 
is obtained. In particular the
line wings agree well, being formed in the deeper layers
of the atmosphere. Small discrepancies are present for the line cores,
in particular for the stronger He\,{\sc i} lines, which originate
at shallower atmospheric depths. And, apparently the
treatment of the He\,{\sc i} forbidden components may be improved in
some cases. 

About the same fit quality can be achieved throughout the
entire parameter space investigated here, with additional minor
complications arising at the hotter end of the $T_\mathrm{eff}$-range,
where weak stellar winds start to impact H$\alpha$ and the He\,{\sc
ii}\,$\lambda$4686\,{\AA} line, see \cite{np07} for examples.

\begin{figure}
\begin{center}
\includegraphics[width=.9\linewidth]{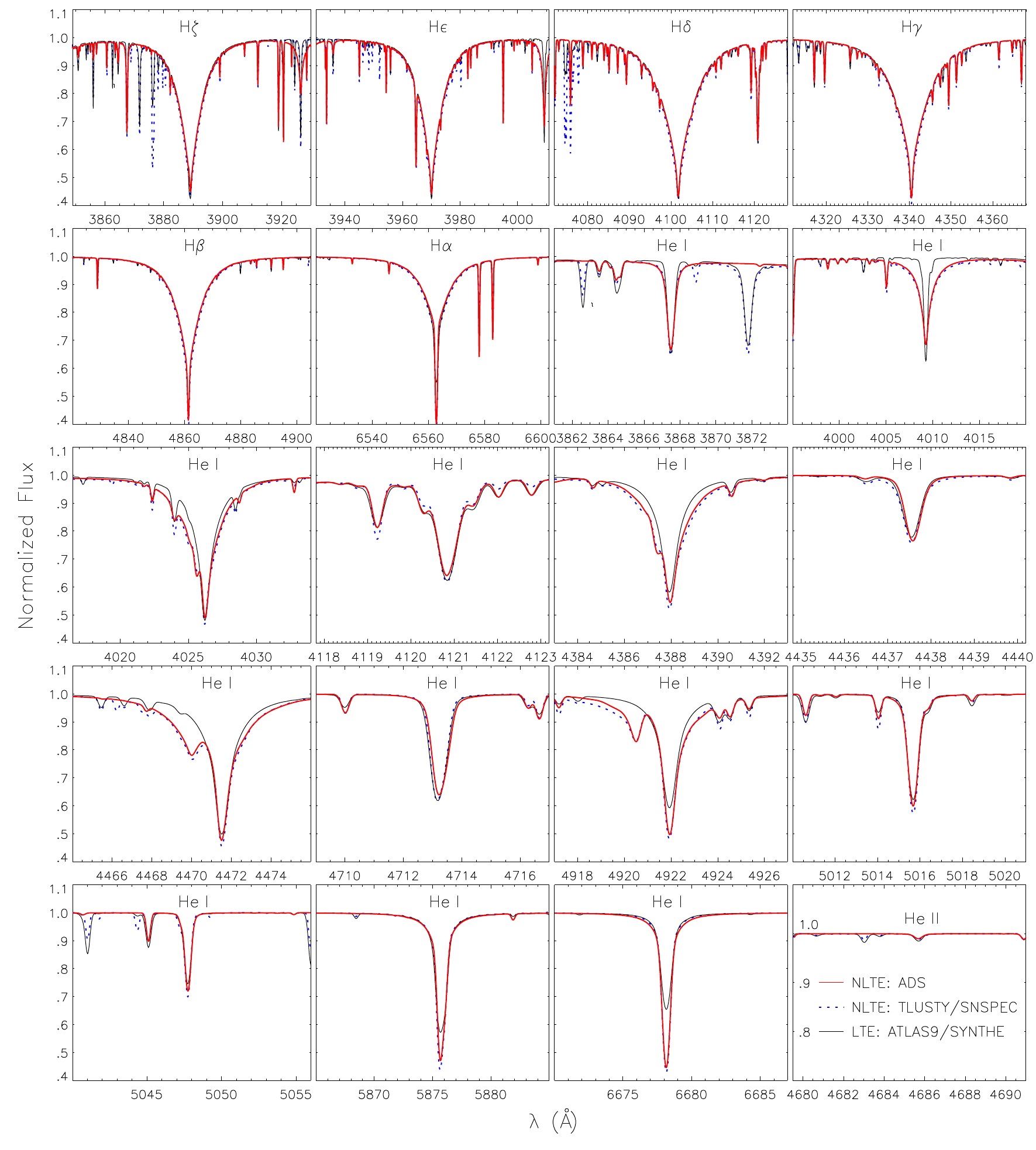}
\end{center}
\vspace{-5mm}
\caption{Same as Fig.~\ref{HHEo}, comparing line profiles computed with {\sc
Ads}, {\sc Tlusty/Synspec} and {\sc Atlas9/Synthe} (line designations
according to the legend).}
\label{HHEt}
\end{figure}

We conclude that the NLTE modelling with {\sc Ads} can reproduce the
observed H\,{\sc i} and He\,{\sc i/ii} lines reliably, except for some
fine details. {\sc Ads} models thus provide a good reference for the
second step of our test, comparison with the {\sc Tlusty/Synspec}
NLTE and {\sc Atlas9/Synthe} LTE models, which is shown in
Fig.~\ref{HHEt}. Both NLTE approaches produce closely resembling
profiles for this test case, and the good agreement (except for some
minor details) holds throughout the entire parameter range considered 
here for the H\,{\sc i}, He\,{\sc ii} and He\,{\sc i} triplet lines.
However, at higher temperatures the He\,{\sc i} singlet lines computed
with {\sc Tlusty/Synspec} become significantly weaker than the {\sc
Ads} profiles (and the observations) -- the well-known singlet-triplet
problem~\cite{nhplm06}. 

Good agreement between the NLTE models and the {\sc Atlas9/Synthe}
profiles is found for the Balmer lines and a few sharp He\,{\sc i}
lines, {\it for this test case}. However, problems with inadequate
line broadening -- Voigt profiles with constant Stark broadening
parameter and no account for forbidden components -- 
are indicated for the majority of the He\,{\sc i}
features (in particular all diffuse ones) computed with {\sc Atlas9/Synthe}.
This can be resolved simply by implementing the appropriate 
broadening tables. More fundamental are the discrepancies in the
depths of lines such as He\,{\sc i}\,$\lambda\lambda$4921, 5875 and
6678\,{\AA}, which are genuine NLTE effects.

The comparison over the entire parameter range 
reveals severe shortcomings of LTE modelling of observed hydrogen and helium
lines. Over the $T_\mathrm{eff}$-range of 22\,000 to 25\,000\,K, 
NLTE effects quickly set in and affect the cores and 
wings of the Balmer lines, and subsequently all helium lines. 
Maximum effects are found for highest $T_\mathrm{eff}$: the Balmer
lines computed in LTE have equivalent widths
about 30\% lower than in NLTE, and up to a factor of more
than 2 for the He\,{\sc i} lines. Surface gravities determinations 
based on Balmer-wing fitting is thus subject to systematic
error, e.g. $\log g$ values are overestimated by up to 0.2\,dex 
from LTE analyses of~H$\gamma$.

\begin{figure}
\begin{center}
\includegraphics[width=.86\linewidth]{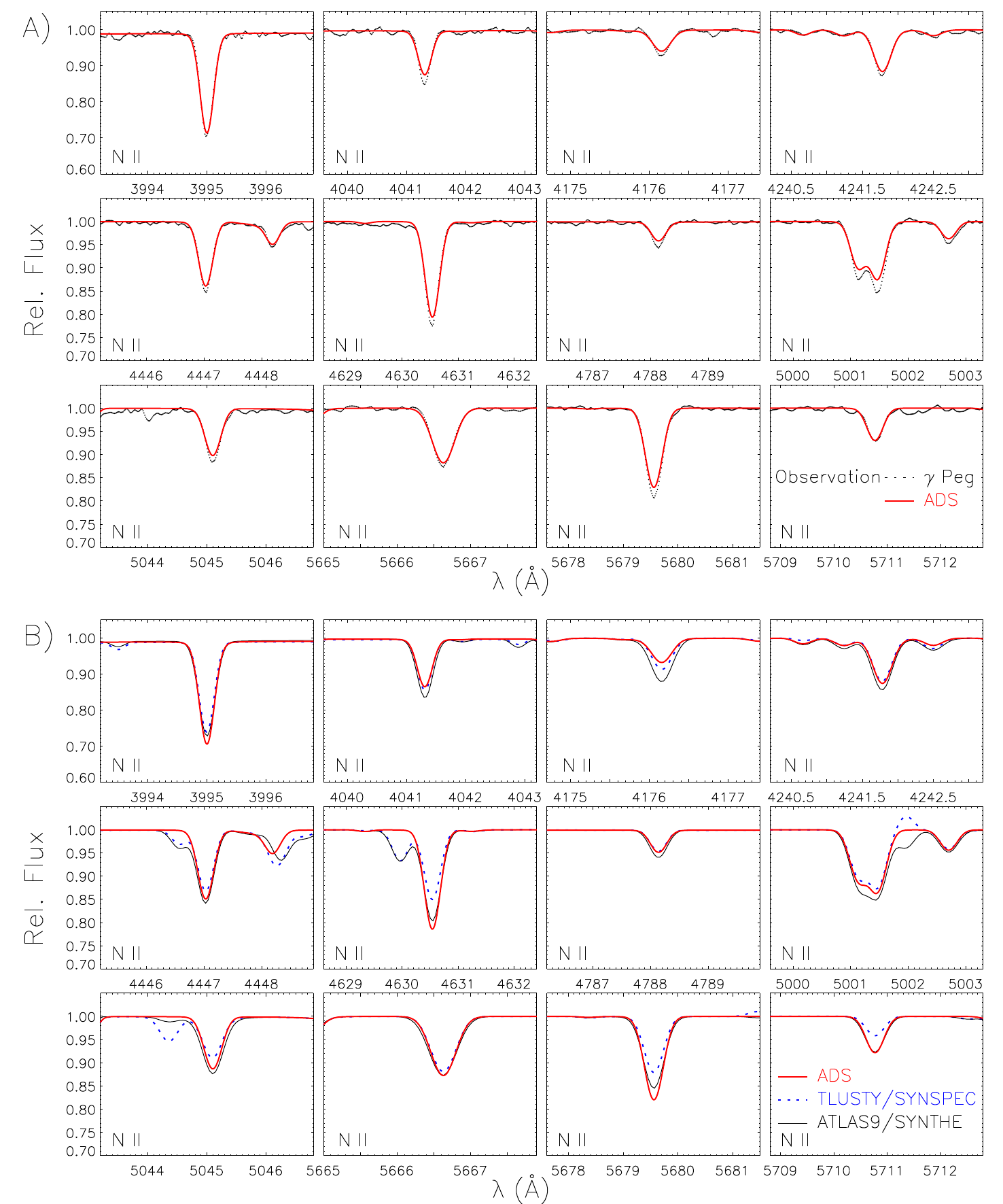}
\end{center}
\vspace{-5mm}
\caption{A):
comparison of observations for N\,{\sc ii} lines in $\gamma$\,Peg with 
our global best-fit {\sc Ads} model. B): comparison
of {\sc Ads}, {\sc Tlusty/Synspec} and {\sc Atlas9/Synthe} models for 
the same lines. See figure legends and the text for details.}
\label{N}
\end{figure}

Overall, LTE modelling of the Balmer wings and selected He\,{\sc i}
lines appears to be largely unbiased by systematic error 
for main sequence stars up to 
$T_\mathrm{eff}$\,$\simeq$\,22\,000\,K. Preference should then be given to
the He\,{\sc i}\,$\lambda\lambda$3867, 4121, 4437, 4713, 5016 and
5048\,{\AA} transitions, which are least affected by NLTE effects.
Either full or hybrid NLTE models have to be used for quantitative
analyses of stars at higher $T_\mathrm{eff}$ when systematic bias is
to be avoided. Our {\sc Ads} approach improves on  the {\sc Tlusty}
OSTAR2002/BSTAR2006 grids by avoiding the He\,{\sc i}
singlet-triplet~problem.\\[1mm]
\noindent \underline{Nitrogen}:
We discuss here only spectral lines of the N\,{\sc ii} ion, which in
contrast to N\,{\sc iii} lines are observable throughout the entire
parameter range. A comparison of important diagnostic
lines in the optical spectrum of $\gamma$\,Peg with our global best-fit 
{\sc Ads} model is shown in Fig.~\ref{N}. The fit quality is excellent, 
in total 40 N\,{\sc ii} lines were analysed to derive the mean
nitrogen abundance. Therefore, the {\sc Ads} model provides a good
reference for the comparison with the other codes, which is shown in
the bottom half of Fig.~\ref{N}.

\begin{figure}
\begin{center}
\includegraphics[width=.9\linewidth]{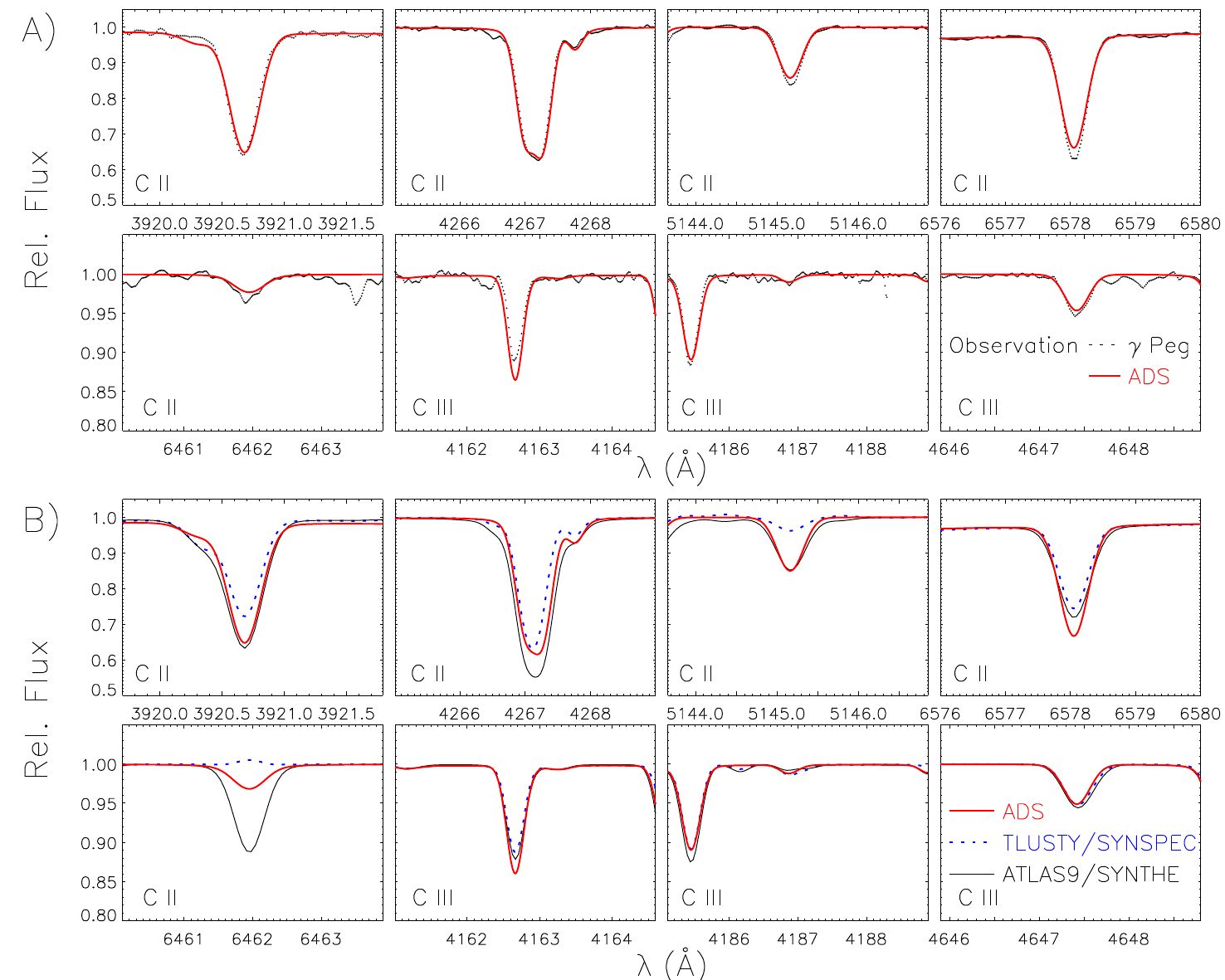}
\end{center}
\vspace{-5mm}
\caption{As Fig.~\ref{N}, for C\,{\sc ii/iii} lines.}
\label{C}
\end{figure}

In many cases good agreement is found between the {\sc Tlusty/Synspec} 
and {\sc Ads} NLTE line profiles. However, a significant number of
differences exists, e.\,g.\,for
N\,{\sc ii}\,$\lambda\lambda$4630, 5045, 5679 and 5711\,{\AA} from the 
examples shown in Fig.~\ref{N}, where the {\sc Tlusty/Synspec} models
are too weak compared to observation. Abundance analyses based on such
features would therefore significantly overestimate the nitrogen
abundance. Both agreement and discrepancies vary
throughout the entire parameter range considered here, so that it is
difficult to be quantify them in brief. In general, the differences
increase towards lower gravities. The reasons for the differences in
the two NLTE approaches result almost entirely from the model
atom implementations. We tested this by adopting the N\,{\sc ii}
model atom used in the {\sc Tlusty} computations for {\sc
Detail/Surface} and could tightly reproduce the {\sc Tlusty/Synspec}
solution.

\begin{figure}
\begin{center}
\includegraphics[width=.86\linewidth]{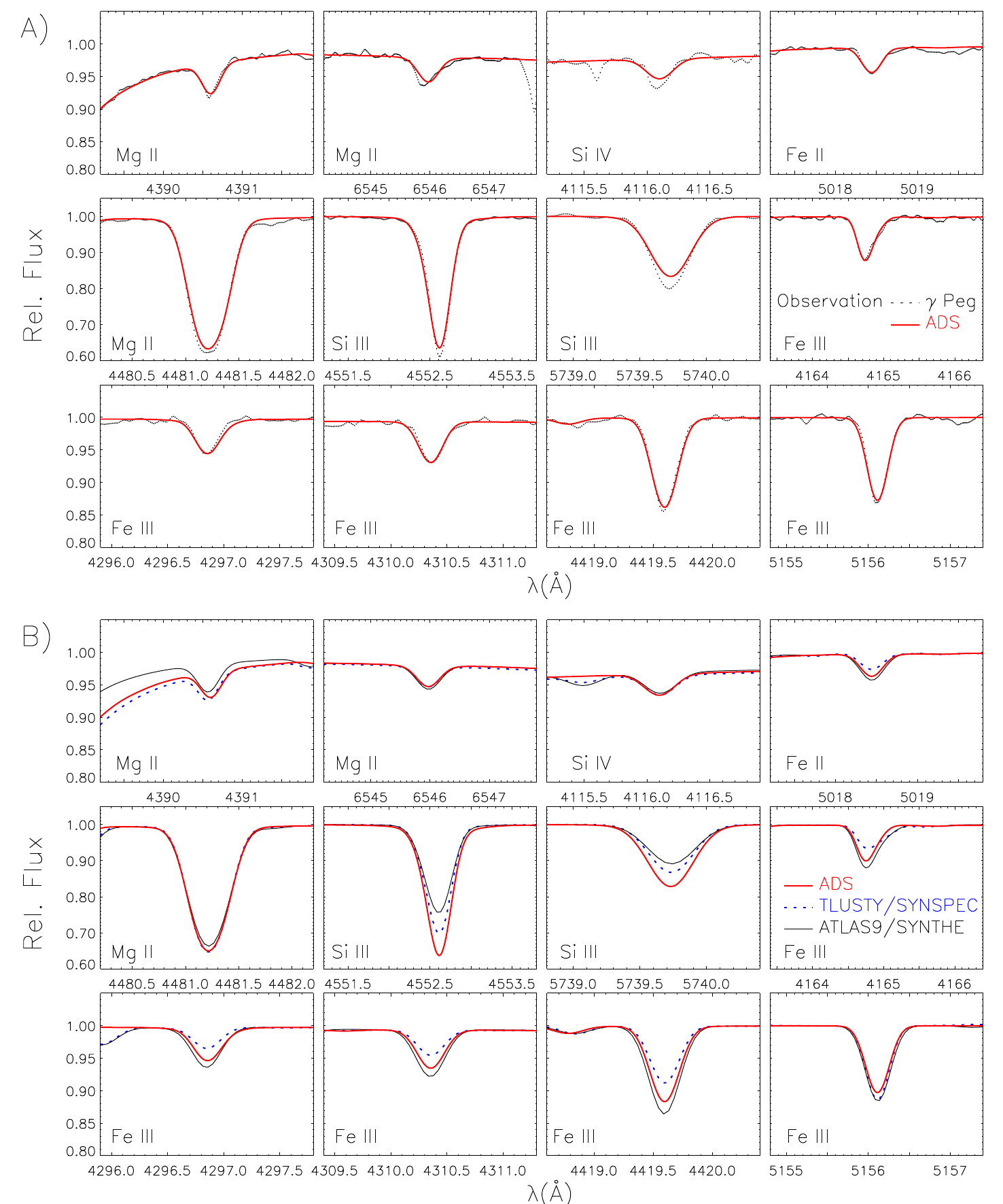}
\end{center}
\vspace{-5mm}
\caption{As Fig.~\ref{N}, for Mg\,{\sc ii}, Si\,{\sc iii/iv} and
Fe\,{\sc ii/iii} lines.}
\label{MGSIFE}
\end{figure}

The comparison of the NLTE {\sc Ads} reference with the LTE {\sc
Atlas9/Synthe} model shows that NLTE effects are often small, 
except for lines such as N\,{\sc ii}\,$\lambda$4176\,{\AA} from the examples.
In fact, {\sc Atlas9/Synthe} profiles are closer to the {\sc Ads}
profiles (and therefore observation) than the {\sc Tlusty/Synspec} profiles 
in most cases. The NLTE effects become more pronounced with decreasing
surface gravities. 
\\[1mm]                                                      
\noindent \underline{Oxygen}: We investigated the behaviour of 
O\,{\sc i} and O\,{\sc ii} lines, with similar results to those 
for nitrogen. Because of this we omit a more detailed description here.
Noteworthy are the overall relatively small NLTE effects, again with the
exception of several multiplets, e.g. the near-IR triplet 
O\,{\sc i}\,$\lambda\lambda$7771--5\,{\AA}.
For these lines, large differences between the two NLTE solutions are seen,
again probably because of the different model atom implementations.\\[1mm]
\noindent \underline{Carbon}: Singly and doubly-ionized carbon were
investigated. Again, the {\sc Ads} model provides a good reference for
further comparison with other models, as it reproduces observation
well, not only for $\gamma$\,Peg (see examples in Fig.~\ref{C}) but
throughout the entire early B-type main sequence~\cite{np08}.

In particular C\,{\sc ii} was known as one of the most challenging
cases for NLTE line formation computations for decades. 
The comparison of {\sc Ads} with the {\sc Tlusty/Synspec} solution 
shows disagreement in practically every line. Usually the {\sc Tlusty/Synspec} 
profiles are too shallow, and in some cases even turn into (unobserved) emission (C\,{\sc ii}\,$\lambda$6462\,{\AA}).
The situation is much better for C\,{\sc iii}, except for the
strong line C\,{\sc iii}\,$\lambda$4162\,{\AA}. 
The problems with
C\,{\sc ii} persist for other parameter combinations. Certainly, the {\sc Tlusty}
model atom for carbon needs to be improved to allow
for meaningful abundance determinations. For the moment, C\,{\sc iii}
should be preferred for abundance work based on the BSTAR2006 and
OSTAR2002 grids.

The comparison with the LTE profiles shows that most of the stronger,
but also several of the weak lines show pronounced NLTE effects,
throughout the entire parameter range. An exception is C\,{\sc
ii}\,$\lambda$5145\,{\AA} (and its other multiplet members), which
stays in LTE for all cases of interest. NLTE effects on the C\,{\sc
iii} lines increase significantly for higher $T_\mathrm{eff}$.\\[1mm]
\noindent \underline{Magnesium, silicon and iron}:
Examples for the comparison of our global best-fit {\sc Ads} model with
observations for important diagnostic Mg\,{\sc ii}, Si\,{\sc iii/iv} 
and Fe\,{\sc ii/iii} lines in $\gamma$\,Peg are shown in Fig.~\ref{MGSIFE}.
The Mg\,{\sc ii} and Fe\,{\sc ii/iii} are reproduced reliably.
We are at present incorporating a new silicon model atom, in particular to
improve the fits to several lines not shown here, but
the old model \cite{bb90} reproduces lines such as those of the 
Si\,{\sc iii}\,$\lambda\lambda$4552--4575\,{\AA} triplet or 
Si\,{\sc iv}\,$\lambda$4116\,{\AA}
reasonably well, providing a valuable diagnostic for
$T_\mathrm{eff}$-determination via the ionization balance.  

Both NLTE solutions, with {\sc Ads} and with {\sc Tlusty/Synspec},
agree well for Mg\,{\sc ii}, but disagree for almost all Si and
Fe lines. This is of concern in the case of silicon as
use of the ionization equilibrium would lead to systematically
different solutions for $T_\mathrm{eff}$ and $\log g$ with the
BSTAR2006 and OSTAR2002 grids, probably
inconsistent with other diagnostics. Moreover, the 
Si\,{\sc iii} triplet is often used to constrain the microturbulence
velocity, which therefore may also become subject to systematic
offsets when based on the {\sc Tlusty} grids. The iron lines are
systematically shallower in the {\sc Tlusty/Synspec} models, implying
the derivation of higher iron abundances in applications. Because of
the r\^{o}le of iron as the main line-opacity source this has further
effects on atmospheric structure computations and therefore also on 
the atmospheric parameter determination. It is highly important to
improve these model atoms for {\sc Tlusty}. 

While the LTE and NLTE solutions for Mg\,{\sc ii} are very similar for the 
parameters of $\gamma$\,Peg (Mg\,{\sc ii}\,$\lambda$4390\,{\AA}
apparently differs because of its location in the wing of an ill-fitted
He\,{\sc i} line), the iron lines are weakly and most of the silicon lines 
strongly NLTE affected. 
LTE modelling of silicon lines in early B-type stars can lead to 
systematic errors in atmospheric parameters when using ionization 
equilibria, and to unreliable elemental abundances. LTE iron abundances
are in general somewhat underestimated.

\section{Conclusions and Recommendations}
We have seen that the data from the Gaia mission will need to be
complemented by ground-based (optical) spectroscopy for early-type
stars in order to put meaningful constraints on atmospheric parameters
and elemental abundances.
Three approaches for the modelling of stellar atmospheres
in the $T_\mathrm{eff}$-range of 15\,000 to 35\,000\,K, corresponding
to spectral types of about B4 to O9, have been tested here: 
full LTE (with {\sc Atlas9/Synthe}), full NLTE ({\sc Tlusty/Synspec}) and 
hybrid NLTE ({\sc Ads}).
The aim of the work was to determine the best means, presently available, to analyse
optical spectra of such early-type stars, which is of course of much broader
relevance than for Gaia-related science alone.

The comparison showed that pure LTE modelling can yield meaningful
results for the low-$T_\mathrm{eff}$ regime, up to about 22\,000\,K on
the main sequence, {\it if the right spectroscopic indicators are employed}. 
An overall good global match between synthetic and observed spectra cannot be
expected to be obtained as many lines show NLTE effects.

Pure NLTE modelling with the available {\sc Tlusty/Synspec} grids is
conceptually superior. However, NLTE atmospheric structures and SEDs turn 
out to be close to LTE in the parameter space considered. 
NLTE line spectra of ions with `simple' atomic structures,
like H\,{\sc i}, He\,{\sc i/ii} (with some restrictions at higher
$T_\mathrm{eff}$) or Mg\,{\sc ii} turn out to be reliable. However,
oversimplified model atoms cause problems for many of the
more complex ions/elements. Blind use of the synthetic spectra
published in the OSTAR2002 and BSTAR2006 grids should therefore be
avoided. {\it The fact is that NLTE computations based on
inappropriate model atoms can potentially introduce larger
systematic errors to quantitative analyses than simple LTE modelling.}
The {\sc Tlusty} model atoms certainly need extensive revision before
a good global match with observations can be obtained with
this package.

Our NLTE line-formation computations with {\sc Detail/Surface} on {\sc
Atlas9} atmospheres combine the advantage of low to moderate CPU
time requirements with realistic model atoms, facilitating a robust
global match of observed spectra to be obtained, see also \cite{nsd11,np11}.
It is reassuring that the same model atoms provide as good a match of
the synthetic spectra to observations also for other kinds of stars, 
with drastically different chemical composition such as subdwarf OB stars
\cite{pne06} or extreme helium stars \cite{pbhj05,pnhj06,ketal10}. 
First applications of our models combined with a sophisticated
analysis methodology promise breakthroughs for
several astrophysical fields, see e.g.~\cite{npi11}.

\end{document}